\documentclass[twocolumn,showpacs,preprintnumbers]{revtex4}

\usepackage{amssymb}
\usepackage{amsmath}
\usepackage{graphicx}
\usepackage{dcolumn}
\usepackage{bm}

\begin{document}

\title{Observation of single- and two-photon beating between independent Raman scattering}
\author{Li-Qing Chen$^{1}$}
\author{Cheng-ling Bian$^{1}$}
\author{Guo-Wan Zhang$^{1}$}
\author{Z. Y. Ou$^{1,2*}$}
\author{Weiping Zhang$^{1\dag}$}
\affiliation{$^{1}$State Key Laboratory of Precision Spectroscopy,
Department of Physics, East China Normal University, Shanghai
200062, P. R. China}
\affiliation{$^{2}$Department of Physics, Indiana University-Purdue University Indianapolis,
402 N. Blackford Street, Indianapolis, IN 46202, USA}

\date{\today }

\begin{abstract}
By using spontaneous Raman processes in the high gain regime,  we produce two independent Raman Stokes fields from an atomic ensemble. Temporal beating is observed between the two directly generated Stokes fields in a single realization. The beat frequency is found to be a result of an AC Stark frequency shift effect. However, due to the spontaneous nature of the process, the phases of the two Stokes fields change from one realization to another so that the beat signal disappears after average over many realizations. On the other hand, the beat signal is recovered in a two-photon correlation measurement, showing a two-photon interference effect. The two-photon beat signal enables us to obtain dephasing information in the Raman process. The dephasing effect is found to depend on the temperature of the atomic medium.
\end{abstract}

\pacs{42.25.Hz,42.25.Kb,42.65.Dr,42.50.Gy}

\maketitle


\section{Introduction}

Interference between independent light sources played an important role in the understanding of the coherence properties of lasers. Interference between two lasers was first demonstrated in 1963 \cite{mag}, shortly after the laser was invented. Its interpretation is based on the coherent states which have a definite phase. This implies that a laser has a definite phase within its long coherence time. Nowadays, it is routine to beat two independent lasers for the measurement of laser frequency stability. However, interference between independent light sources is not limited to coherent states. As a matter of fact, the first experiment that demonstrated the interference between independent light sources is the beat from two fluorescent sources \cite{for}. More recently, Kuo et al \cite{kuo} demonstrated in 1993 the interference effect between two pulsed thermal sources in two independent Raman amplification processes. But thermal sources of light have a completely random phase relation.  Its interpretation is that the Raman amplification process (stimulated emission) preserves the initial phase from vacuum induced spontaneous emission but different pulses start from different spontaneous emission and thus have random phases. Theoretical study even suggested interference between two groups of photons in number states whose phase is completely undetermined although each quantum realization have a definite but random phase \cite{yoo,ou02}. Further study showed that this interference effect is a result of multi-photon interference and the concept of coherence has something to do with indistinguishability among photons involved \cite{ou08}.

Furthermore, higher-order interference such as two-photon interference may not vanish even after the average over the phase fluctuations. Indeed, light from a laser is more accurately described by a coherent state with a diffused phase. So, interference fringe will be washed out in the long-term or multiple exposures in one detector but may appear in the two-photon correlation measurement with two detectors \cite{pfla,pflb}. Similar situation occurs with thermal light \cite{man83,kuo}.
The difference between a coherent state and a thermal state is simply the amplitude fluctuations, which can be reflected in higher order measurement of fourth-order (two-photon) interference: the visibility of the two-photon interference is 1/2 for a coherent state with phase diffusion but is 1/3 for a thermal state \cite{pfla,pflb,man83,kuo}.

In all previous studies of interference between independent sources, it is crucial that in a single pulse generation process, the phases are fixed even though they may fluctuate from one pulse to another. This will ensure the formation of an interference fringe. However, because of the complexity of the systems involved, dephasing in the light generation process does occur. In other words, the phase may change randomly even during the single pulse generation process. For Raman scattering process, light waves (the incoming pump and scattered Stokes) are coupled through phonons in solid or atomic spin wave in atomic medium. Dephasing usually occurs more often in phonons or atomic spin wave than in optical waves. Coherent atomic spin wave plays a crucial role in the scheme of quantum memory realized in atomic medium \cite{fle,hau,phi}. Dephasing time in atomic spin wave thus determines the quantum storage time.

However,  such a dephasing process cannot be observed in the interference fringe pattern from a single detector and may not appear even in the spatial fringe pattern in two-photon interference \cite{kuo}.
Since the dephasing process involves time evolution of the phases, the best way to characterize it is by a time-resolved two-photon correlation measurement, which compares the intensities, or the interference fringes in our case here, at two different times. Furthermore, time-resolved two-photon correlation measurement is also a tool for the direct observation of two-photon beating effect which is an indication of two-photon frequency entanglement \cite{ou88a}.    On the other hand, time-resolved two-photon correlation measurement requires that the detector's response be faster than the fluctuations of the fields as well as the beat signal, which is also the requirement for the observation of two-photon interference effect between two independent fields \cite{ou88b}. So far, two-photon frequency entanglement has only been confirmed indirectly via spatial beating effect because of this requirement \cite{ou88a,li09}.

Here in this paper,  we report on an interference experiment between two independent light sources, in which we observed the dephasing effect in the light generation process by a time resolved two-photon correlation measurement. In the experiment, we observe a beat signal between independent  Stokes fields generated in two separate Raman scattering processes in an atomic vapor cell. Although the beat signal disappears after average over many pulses, indicating a random phase relation between the two fields, it is recovered when we measure the intensity correlation, thus showing a two-photon interference effect between independent sources. The observed visibility in the two-photon beat signal decays as the time delay increases, indicating the existence of dephasing due to the atomic motion in the process.

\section{Experimental set up and observation of beating between independent sources}

The conceptual diagram and the energy levels of atom and light fields are shown in Fig.1(a,b). The experimental layout is shown in Fig.1(c) together with the time sequence of the light pulses. A Toptica-DL100 semiconductor laser, modulated by an acoustic optic modulator (AOM), is used to provide linearly polarized light pulses at 795nm with duration 20$\mu$s. The laser beam, now called the ``write" beam, is split into two beams ($W_{1,2}$) which are sent
in parallel through a cylindrical Pyrex cell (length and diameter were
75mm and 19mm, respectively) for Raman scattering. The cell is filled with isotopically enriched Rb-87 without buffer
gas and is mounted inside a four-layer $\mu$-magnetic shielding to reduce stray magnetic fields and can be heated up to 95$^{\circ}$C using a bi-filar
resistive heater. The spot sizes of the two write beams   are 0.45mm ($W_1$)
and 0.63mm($W_2$) in diameter, respectively. As shown in the energy diagram in Fig.1(b), states
$|1\rangle =|5S^{1/2},F=1\rangle $ and $|2\rangle =|5S^{1/2},F=2\rangle$ are the hyperfine splitting of the
ground state of the Rubidium atom, $|3\rangle =|5P^{1/2},F=2\rangle$ and $|4\rangle =|5P^{3/2},F=1\rangle$ are the excited states.
The optical pumping pulses ($P$)
are applied before the write pulses ($W_{1,2}$) to prepare the atoms in the state
$|1\rangle =|5S^{1/2},F=1\rangle $ (see inset of Fig.1 for timing sequence). The two independent Stokes fields that are generated by the two separate Raman processes
are coupled into a single-model fiber for superposition  and then detected by a photodiode.

\begin{figure}[tbp]
\centerline{\includegraphics[angle=-90,bb=0 0 550
680,scale=0.45]{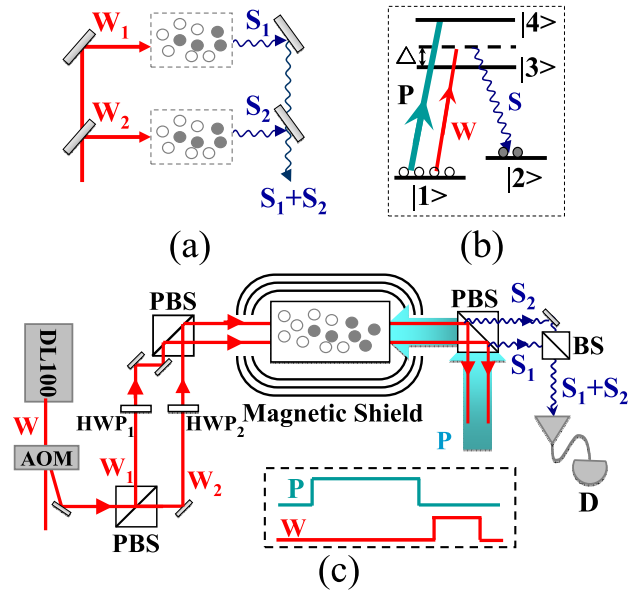}}
\caption{(a) The conceptual diagram for the beating between two independent Stokes fields (b) Diagram for the atomic energy level and light frequencies. $W_{1,2}$: the Raman write fields; $S_{1,2}$: the generated Stokes fields. (c) Experimental layout. Inset: timing sequence.} \label{fig1}
\end{figure}

\begin{figure}[tbp]
\centerline{\includegraphics[scale=0.5,angle=0]{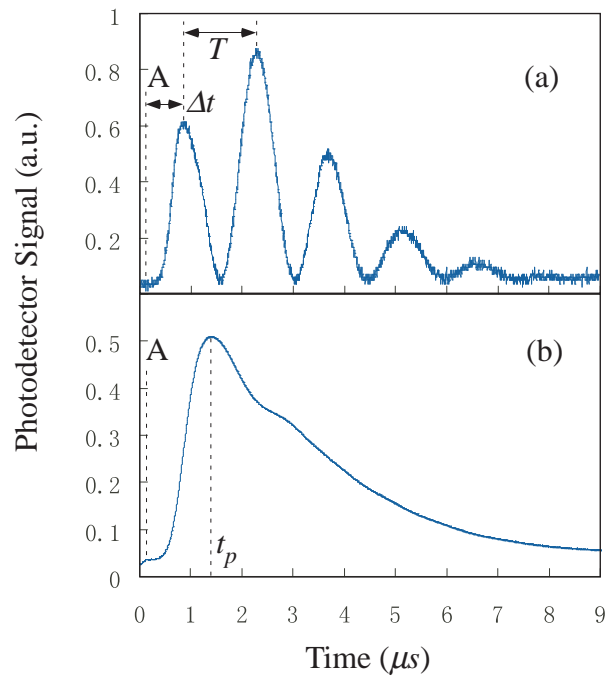}}
\caption{(a) Single-pulse beat signal from the photodetector. (b) Average over 2000 pulses.}
\label{fig2}
\end{figure}

Fig.2 shows some typical results from the photodiode. Fig.2a is for a single pulse, showing an interference fringe pattern in the form of a frequency beating in time. This clearly demonstrates the good coherence between the two independent sources of light in Raman scattering process. Fig.2b is an average over 2000 pulses, washing out the beat signal.

Before we discuss other issues, let us find out the origin of this beat signal. From Fig.2a, we find a typical period is about 1.3 $\mu s$ which corresponds to a beat frequency of $\Delta \nu = 0.77 MHz$. Because of the two-photon resonance condition for Raman gain, we would expect the same frequency for the Stokes fields since the two writing fields have the same frequency. The only difference in the conditions for the generation of the two Stokes fields is the powers of the writing fields. This points to the AC Stark shift as the possible cause for the frequency difference in the two Stokes. It is well-known \cite{AC} that when atoms are illuminated by a beam of light, their energy levels are shifted up by the amount of $\Delta E = |\Omega|^2/\Delta$ with $\Omega = dA_W/\hbar$ as the Rabi frequency and $\Delta$ as the detuning ($d$ is the atomic dipole moment and $A_W$ is the amplitude of the write field). So the frequency shift $\Delta f$ is proportional to the power $P_W (\propto |A_W|^2)$ of the write beam: $\Delta f \propto P_W$. For our experiment, the two-photon resonance condition leads to the beat frequency between the two Stokes fields:
\begin{eqnarray}
\Delta \nu = \kappa_1 P_{W1} - \kappa_2 P_{W2}, \label{beat}
\end{eqnarray}
where $\kappa_1,\kappa_2$ are some proportional constants. Note we choose different proportional constants for the two beams because the electric field strength also depends on the geometry of the beam such as the beam waist.

To confirm the dependence in Eq.(\ref{beat}), we measured the beat frequency as we change via $HWP_1$ the power of one of the writing beams ($W_1$) while fix the other at $P_{W2} = 0.24 mW$. Fig.3 plots the beat frequency as a function of the power of the writing beam $W_1$. It can be seen that the data follows very well the linear dependence given in Eq.(\ref{beat}).

\begin{figure}[tbp]
\centerline{\includegraphics[scale=0.5,angle=0]{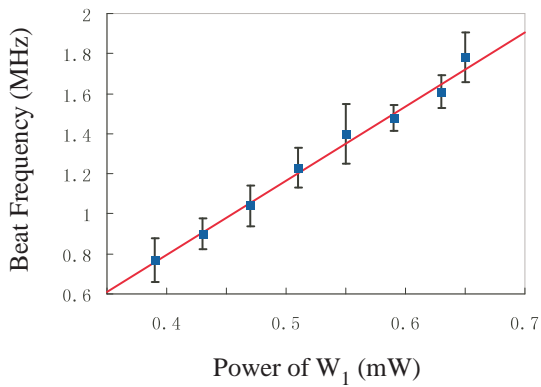}}
\caption{The beat frequency as a function of the power of the write beam ($W_1$). The solid line is a linear fit.}
\label{fig3}
\end{figure}

\section{Random phases and two-photon beating effect}

Now we turn to the important issue of (de)coherence between the two Raman scattering processes. Because of the spontaneous nature of the Raman scattering process, the two Stokes fields are initiated from two different vacuo and the phases of the two Stokes fields are not correlated. Therefore, the two processes can be regarded as independent of each other. So, after an average over many pulses, the beat signal is gone, as seen in Fig.2b. To further confirm the random phase relationship, we directly measure the phase from the beat signal by measuring the position ($\Delta t$) of the first maximum relative to a fixed reference point, the switch-on point of the write pulse (point $A$ in Fig.2a), and comparing it with the period of the beating ($T$):
 $\Delta\varphi_{S} = 2\pi \times \Delta t/T$. Here $T$ are the average period of the beat signals.
Fig.4a plots the extracted phases for a sequence of 5000 pulses. A probability distribution is calculated from the data and is shown in Fig.4b. As can be seen, it is a random phase distribution.

\begin{figure}[tbp]
\centerline{\includegraphics[scale=0.4,angle=0]{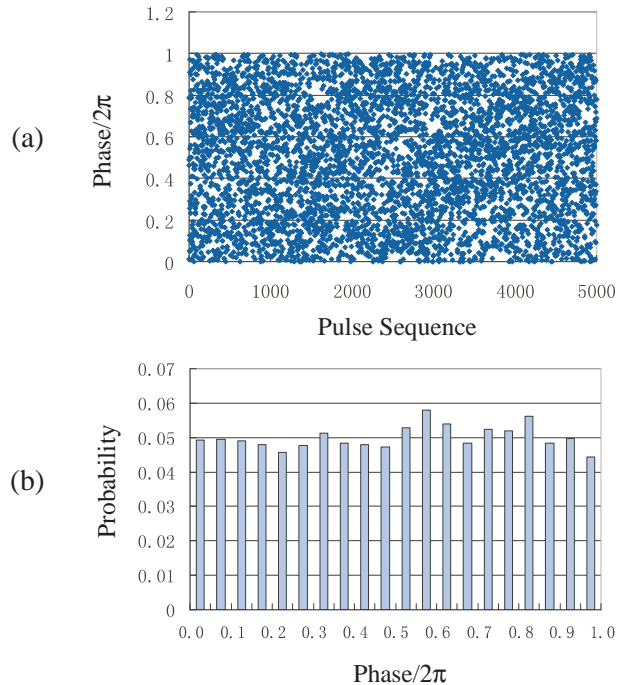}}
\caption{(a) Extracted phases for 5000 pulses and (b) Probability distribution for the extracted phase.}
\label{fig4}
\end{figure}

Because of the random phase, after the average over many  pulses, the interference effect vanishes in the signal of a single detector. However, as we discussed earlier, it can appear in a higher order quantity such as  the two-time intensity correlation function $\Gamma^{(2)}(\tau) \equiv \langle I(t)I(t+\tau)\rangle$, where the average is over many pulses. Indeed, if we plot the normalized intensity correlation function $g^{(2)}(\tau) \equiv \Gamma^{(2)}(\tau)/\langle I(t)\rangle\langle I(t+\tau)\rangle$ as a function of the time delay $\tau$, the beat signal appears again, as shown in Fig.5 as the blue line.

The data in Fig.2a is a single trace obtained from one pulse. The intensity correlation function $\Gamma^{(2)}(\tau)$ and the intensity average $\langle I(t)\rangle, \langle I(t+\tau)\rangle$ are evaluated by averaging the data in Fig.2 over the traces from many pulses. Fig.2b shows the result of $\langle I(t)\rangle$. But because $\langle I(t)\rangle$ depends on time $t$, we need to choose the initial time $t$ for the evaluation of $\Gamma^{(2)}(\tau) \equiv \langle I(t)I(t+\tau)\rangle$. In Fig.5, we choose $t=t_p$ at the peak of $\langle I(t)\rangle$ as the zero delay point.
\begin{figure}[tbp]
\centerline{\includegraphics[scale=0.5,angle=0]{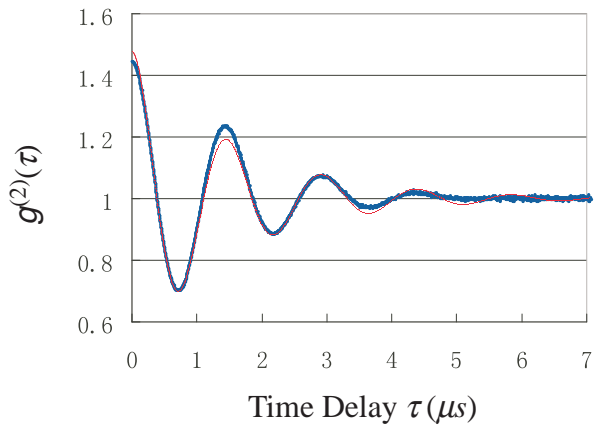}}
\caption{Normalized intensity correlation function $g^{(2)}(\tau)$ as function of time delay $\tau$. Blue line: experimental data; red line: best fitted curve to Eq.(\ref{gde}) with $V=0.47, \gamma=0.63 \mu s^{-1}, \Delta \nu = 0.68 MHz$. }
\label{fig5}
\end{figure}

The two-photon beating effect can be easily explained with a simple model. Consider that for one write pulse, the initial spontaneous
emission of the Stokes field carries an arbitrary phase but the stimulated emission afterwards preserves this phase. So the observed Stokes field in this process will carry
a definite but arbitrary phase. On the other hand, different light intensity
causes different AC Stark shift. This is the origin of the beat signal detected in a single optical detector. From any optics textbook, we can find the beat signal as
\begin{eqnarray}
I(t) = U(t)[
I_1 + I_2 + 2\sqrt{I_1I_2} \cos (2\pi\Delta \nu t + \Delta\varphi)],\label{I}
\end{eqnarray}
where we assumed that the two Stokes fields have the same temporal profile $U(t)$ and is given in the average in Fig.2b. $I_{1,2}$ are the intensity of the two Stokes fields. $\Delta\nu$ is the frequency difference due to AC Stark shift and $\Delta\varphi$ is the random phase difference due to spontaneous nature of the Raman scattering process.

We now calculate the two-time correlation function $\Gamma^{(2)}(\tau) \equiv \langle I(t)I(t+\tau)\rangle$ and $\langle I(t)\rangle$ from Eq.(\ref{I}). Here the average is over the random phase difference $\Delta\varphi$ because it varies from one pulse to another. We obtain
\begin{eqnarray}
&&\Gamma^{(2)}(\tau)=
U(t)U(t+\tau) [(I_1+I_2)^2 + 2I_1I_2 \cos 2\pi\Delta \nu \tau],\cr
&&\langle I(t)\rangle = U(t)
(I_1 + I_2).
\end{eqnarray}
So the normalized two-time correlation function:
\begin{eqnarray}
g^{(2)}(\tau) &\equiv& \Gamma^{(2)}(\tau)/\langle I(t)\rangle\langle I(t+\tau)\rangle
\cr &=& 1+ V\cos 2\pi\Delta \nu \tau ,
\end{eqnarray}
with $V= 2I_1I_2/(I_1+I_2)$ as the visibility. The visibility is 0.5 when $I_1=I_2$ and is independent of the delay.

\section{Dephasing effect}

However, Fig.5 clearly shows the drop of the beat visibility as the delay increases. So we need to modify the simple model above by introducing dephasing. The physics of this dephasing is the random atomic motion in a hot atomic medium. Because of this, the generated spin wave  no longer keeps the same initial phase of the spontaneous emission, which in turn leads to the change of the phase of the Stokes field in the generation process. Thus, even for a single pulse, the phase difference $\Delta\varphi$ is no longer a constant but a function of time: $\Delta\varphi= \Delta\varphi (t)$. So we have
\begin{eqnarray}
\Gamma_{de}^{(2)}(\tau) &=&
U(t)U(t+\tau) [(I_1+I_2)^2 \cr && + I_1I_2 \{e^{ 2i\pi\Delta \nu \tau}\langle e^{i[\Delta\varphi(t+\tau)-\Delta\varphi(t)]}\rangle + c.c\}].\label{Gdp}
\end{eqnarray}
The dephasing can usually be characterized by a decay constant $\gamma$ as
\begin{eqnarray}
\langle e^{i[\Delta\varphi(t+\tau)-\Delta\varphi(t)]}\rangle = e^{-\gamma \tau}.
\end{eqnarray}
So Eq.(\ref{Gdp}) becomes
\begin{eqnarray}
\Gamma_{de}^{(2)}(\tau) &=&
U(t)U(t+\tau) [(I_1+I_2)^2 \cr && + 2I_1I_2 e^{-\gamma \tau} \cos 2\pi\Delta \nu \tau].\label{Gdp2}
\end{eqnarray}
and the normalized intensity correlation function is
\begin{eqnarray}
g_{de}^{(2)}(\tau) = 1+ V e^{-\gamma \tau} \cos 2\pi\Delta \nu \tau.\label{gde}
\end{eqnarray}
In Fig.5, the red thin line is the best-fitted curve to Eq.(\ref{gde}) with $V=0.47, \gamma=0.63 \mu s^{-1}, \Delta \nu = 0.68 MHz$. It can be seen that there is a reasonably good fit between the blue experimental data and the red theoretical curve. Since the dephasing effect is due to atomic motion, it should depends on the speed at which the atoms move. For an atomic cell, the average atomic speed is determined by the temperature of the cell. In Fig.6, we plot the decay constant $\gamma$ as a function of the cell temperature. As expected, the dephasing rate is larger for higher cell temperature.

\begin{figure}[tbp]
\centerline{\includegraphics[scale=0.55,angle=0]{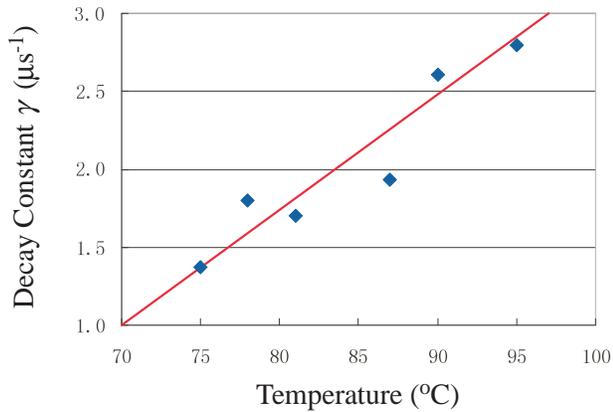}}
\caption{Decay constant as a function of the temperature of the cell. Solid line is for observation guidance.}
\label{fig6}
\end{figure}

From Fig.5 we find the peak visibility is near 50\% ($V=0.47)$. This is consistent with the situation when the two interfering fields are in coherent states. However, it was proven that the Stokes field from Raman scattering is normally in a thermal state \cite{li}, which should lead to a visibility of 1/3 in two-photon interference \cite{man83}. On the other hand,  because of the limited number of atoms available for Raman scattering, we may have used up all the atoms and the Raman amplification process reaches a saturated stage, in which the intensity fluctuation is of the nature of a coherent state. This explains well the observed visibility of $V=0.47$ but not 1/3.

\section{Summary}

In summary, we demonstrated one-photon and two-photon beating between two independent Raman Stokes fields. From the two-photon beat signal we are able to observe a dephasing process during the Raman amplification process. This dephasing effect is due to atomic motion in the atomic cell and will influence the coherence of the atomic spin waves that are crucial in the application of atomic memory.

\begin{acknowledgments}

This work is supported by the National Natural Science Foundation of China
under Grant Nos. 10828408 and 10588402, the National Basic Research
Program of China (973 Program) under Grant No. 2006CB921104,  the Program of Shanghai Subject Chief Scientist under Grant No. 08XD14017, Shanghai Leading Academic Discipline Project under Grant No. B480, Shanghai outstanding young teacher fund.
\newline
Email:$^*$zou@iupui.edu; $^\dag$wpzhang@phy.ecnu.edu.cn
\end{acknowledgments}


\end{document}